\documentclass[letter,twocolumn]{jpsj3}
\usepackage{txfonts}
\usepackage{cuted}
\usepackage{flushend}
\title{Lorentz Covariance of Dirac Electrons in Solids: 
\\
Dielectric and Diamagnetic Properties}

\author{Hideaki Maebashi$^1$,  Masao Ogata$^1$, and Hidetoshi Fukuyama$^2$}
\inst{$^1$Department of Physics, University of Tokyo, Bunkyo, Tokyo 113-0033, Japan \\
$^2$Department of Applied Physics, Tokyo University of Science, 
Shinjuku, Tokyo 162-8601, Japan} 

\abst{We study the electrodynamics of Dirac electrons in solids (e.g., bismuth) 
by comparing it with quantum electrodynamics (QED).
It is shown that Lorentz covariance associated with the Dirac electrons in solids 
results in a remarkable correlation between the dielectric and diamagnetic properties, 
leading to a significant enhancement in the permittivity directly linked to the well-known 
phenomenon of large diamagnetism.}

\begin{document}
\maketitle

The Dirac equation is the cornerstone of relativistic quantum mechanics, 
and it was originally derived by Dirac requiring the electron wave equation 
linear in time-derivative to be Lorentz--covariant~\cite{Dirac28}. 
As pointed out by Wolff~\cite{Wolff64}, an essentially equivalent equation describes 
the motion of nonrelativistic electrons in narrow-gap systems with strong 
spin--orbit coupling such as bismuth~\cite{Fuseya15}.  
The Dirac equation in narrow-gap systems is invariant under a Lorentz 
transformation when the speed of light $c$ is replaced by 
an effective speed of light 
${c^*} \!=\! \sqrt{E_{\rm G}/2{m^*}}$ with $E_{\rm G}$ and ${m^*}$ as 
the band gap and effective electron mass, respectively.
As a result, not in the original sense but another type of Lorentz covariance specified 
by ${c^*}$ emerges for the Dirac electrons in solids.
The Dirac electron system is not just interesting in itself but also 
provides a platform to study topological insulators~\cite{HasanKane10,Qi11} 
and exotic magnetoelectric effects~\cite{Burkov15}.

One of the most interesting phenomena in the Dirac electrons of solids is large 
diamagnetism which has been experimentally known for many years, e.g. in Bi,
and the magnitude of diamagnetism is at a maximum  
when the chemical potential is located in the band gap~\cite{Shoenberg36,Wehrli68}. 
This has been theoretically explained by an interband effect of the magnetic 
field~\cite{Fukuyama70}.
Thus, it is distinct from Landau diamagnetism which results from the Landau quantization 
of electron orbital motion in metals~\cite{Landau30}. 
Based on the Luttinger--Kohn representation~\cite{Luttinger55}, which is equivalent to 
the standard Bloch representation when linked by a unitary transformation, a general formula 
for the uniform and static orbital susceptibility has previously been 
established~\cite{Fukuyama71}. 
With this formula, the diamagnetic properties of Dirac electron systems and related materials have been intensively studied~\cite{McClure74,Fukuyama07,Koshino07,Kobayashi08,Fuseya14,Ogata15,Ogata16-1,Ogata16-2,Ogata17}. 
However, the dielectric properties and electrodynamics of Dirac electrons in solids
have not yet attracted much attention~\cite{Boyle60,Edelman75}. 
Apparently, the electrodynamics of Dirac electrons in narrow-gap systems can be 
taken as a counterpart of quantum 
electrodynamics~\cite{Peskin95,Schwinger58,Tomonaga48,Ward50} (QED) in solids.
In Table~\ref{tab:1}, we present a correspondence table for Dirac electrons 
in bismuth and QED.
In particular, the zero-temperature insulator with the greatest diamagnetism 
corresponds to the vacuum in QED.

In this letter, we report our theoretical results on electric susceptibility 
$\chi_{\rm e}(q,\omega) = \varepsilon_{\rm r} (q, \omega) - 1$ 
and magnetic susceptibility $\chi_{\rm m}(q,\omega) = 1 - 1/\mu_{\rm r} (q, \omega)$ 
with a magnitude $q$ of a wave vector and frequency $\omega$
of Dirac electrons in solids, 
where $\varepsilon_{\rm r} (q, \omega)$ and $\mu_{\rm r} (q, \omega)$ 
are the relative permittivity and permeability, respectively\cite{Note1}. 
We find the relationship between the susceptibilities 
to be $\chi_{\rm e}(q,\omega) = - (c^2/{c^*}^2)\chi_{\rm m}(q,\omega)$ 
for the zero-temperature insulator, 
which originates from Lorentz covariance of the Dirac equation and 
can be considered as the nature of the vacuum in QED that is
realized in solids. 
With this relationship and the explicit evaluation of the charge renormalization factor 
$Z_3 \!\equiv\! 1/\varepsilon_{\rm r} (0, 0)$ in solids, 
we show a significant enhancement 
of the permittivity directly linked to the large diamagnetism. 
\begin{table}[t]
\caption{Correspondence table for Dirac electrons in bismuth and QED. 
$e_0$ is the bare electric charge 
and $Z_3$ is a factor associated with the charge renormalization.
$m$ and $e$ are the electron mass and elementary charge, respectively. 
See more details in the text.}
\label{tab:1} 
\centering
\begingroup
\renewcommand{\arraystretch}{1.2}
\begin{tabular}{lcc}
\hline
& \hspace{3.5em} Bismuth\hspace{3.5em}   & \hspace{3em} QED\hspace{3em}   
\\
\hline
\hspace{1.23em}Energy scale\hspace{0.5em} & $\sim\!10^{-2}$ eV  & $\sim\!10^{6}$ eV  \\
\hspace{1.23em}${c^*}/c$ & $\sim\!10^{-3}$ & $1$  \\
\hspace{1.23em}${m^*}/m$ & $\sim\!10^{-2}$ & $1$  \\
\hspace{1.23em}$\,e_0/e$ & 1 & $1/\sqrt{Z_3}$ \\
\hspace{1.23em}Permittivity & $\varepsilon_0 \varepsilon_{\rm r}(q, \omega)$ 
& $\varepsilon_0 Z_3 \varepsilon_{\rm r}(q, \omega)$  \\
\hspace{1.23em}Permeability & $\mu_0 \mu_{\rm r}(q, \omega)$ & 
$\mu_0 Z_3^{-1} \! \mu_{\rm r}(q, \omega)$  \\
\hline
\end{tabular}
\endgroup
\end{table}

First, we note the nature of the vacuum in QED~\cite{Peskin95}. 
The permittivity $\varepsilon_0$ and permeability $\mu_0$ of the classical vacuum 
are constants and related to each other by $\varepsilon_0 \mu_0 = c^{-2}$ 
due to the Lorentz covariance of Maxwell's equations~\cite{Jackson99}. 
However, in QED, the vacuum permittivity and permeability are not constants
but depend on $q$ and $\omega$ as $\varepsilon_0 \varepsilon_{\rm r} (q, \omega)$ and 
$\mu_0 \mu_{\rm r} (q, \omega)$, respectively. 
In particular, $\varepsilon_{\rm r} (q, \omega)$ describes the vacuum polarization 
caused by the dynamics of virtually excited particle--antiparticle pairs~\cite{Levine97}. 
Even in the case of the polarized vacuum, 
the Lorentz covariance of the Dirac equation 
makes a desired correlation between the electric and magnetic properties of the vacuum 
as $\varepsilon_{\rm r} (q, \omega) \mu_{\rm r} (q, \omega) 
\!=\! 1$~\cite{Weldon82}.
In the uniform and static limit of $q\!=\!\omega\!=\!0$, 
the previous equation reduces to 
$Z_3 \!\equiv\! 1/\varepsilon_{\rm r} (0, 0) \!=\! \mu_{\rm r} (0, 0)$. 
We can then renormalize $\varepsilon_{\rm r} (q, \omega)$ and $\mu_{\rm r} (q, \omega)$ 
as $\varepsilon_{\rm r}^* (q, \omega) \!=\! Z_3 \varepsilon_{\rm r} (q, \omega)$ and 
$\mu_{\rm r}^* (q, \omega) \!=\! Z_3^{-1}\mu_{\rm r} (q, \omega)$, respectively,  
such that $\varepsilon_{\rm r}^* (q, \omega)$ and $\mu_{\rm r}^* (q, \omega)$ 
are equal to $1$ in the $q\!=\!\omega\!=\!0$ limit.
Correspondingly, the bare electric charge $e_0$ is assumed to be renormalized as 
$e_0^* \!=\! \sqrt{Z_3} e_0$ in QED~\cite{Note2}, 
and the physically observable elementary charge, permittivity, and permeability  
are identified as $e_0^*$, $\varepsilon_0\varepsilon_{\rm r}^*(q, \omega)$, 
and $\mu_0\mu_{\rm r}^*(q, \omega)$, respectively, in QED~\cite{Peskin95,Weldon82}. 
This renormalization procedure is also summarized in Table~\ref{tab:1}. 
In QED, the value of $Z_3$ cannot be determined 
because it is renormalized into the elementary charge $e$~\cite{Tomonaga48}. 
It is important to note that this is not the case in solids.

We begin by introducing the Dirac Hamiltonian in solids, which is effectively 
identical to the Wolff Hamiltonian that describes 
low-energy electron excitations in narrow-gap systems~\cite{Wolff64}.  
The Dirac Hamiltonian is given in its second quantized form as
\begin{align}
H = \sum_{\boldsymbol k} {\bar \psi}_{\boldsymbol k}
\left[ \hbar {c^*} k_i \gamma^i + {m^*} c^{*2}  \right] \psi_{\boldsymbol k} 
\label{eq: 1}
\end{align}
with ${\bar \psi}_{\boldsymbol k}\!=\!\psi_{\boldsymbol k}^{\dagger} \gamma^0$ 
and where ${\boldsymbol k} \!=\! (k_1,k_2,k_3)$ is a wave vector, 
$\gamma^0$, $\gamma^1$, $\gamma^2$, and $\gamma^3$ 
are the gamma matrices, and the repeated Roman indexes $i\!=\!1,2,3$ are to be summed. 
Under a canonical transformation,
the four components of $\psi_{\boldsymbol k}$ correspond to the conduction and valence band 
electrons with a spin degeneracy in 
the Luttinger--Kohn representation~\cite{Wolff64,Luttinger55,Cohen60}. 
In the Dirac Hamiltonian, Eq.~(\ref{eq: 1}), 
anisotropy of the effective mass, which has been considered in the Wolff Hamiltonian, is neglected.
In a forthcoming paper, we plan to investigate the effects of anisotropy 
in comparison between theory and experiment for the permittivity.

The coupling of Dirac electrons with an electromagnetic field is obtained by the 
gauge principle with the electromagnetic scalar and vector potentials as  
$\phi_{\boldsymbol q} (\omega)$ and ${\boldsymbol a}_{\boldsymbol q} (\omega)$, 
respectively~\cite{Luttinger55},
resulting in an additional time-dependent Hamiltonian
$H^{\prime}(t) \!=\! - e_0 \sum_{{\boldsymbol q}} J^{\,\mu}({\boldsymbol q})
A_{\mu} (- {\boldsymbol q},\omega)  e^{- i \omega t}$, 
where the repeated Greek indexes $\mu\!=\!0,1,2,3$ are to be summed. 
With the use of ${c^*}$ instead of the conventional use of $c$, 
we define a four-current and an electromagnetic four-potential as 
${J}^{\mu}({\boldsymbol q}) \!=\! ({c^*}\! \rho_{\boldsymbol q}, 
{\boldsymbol j}_{\boldsymbol q} ) \!=\!  {c^*} \sum_{\boldsymbol k} 
{\bar \psi}_{{\boldsymbol k}+{\boldsymbol q}} \gamma^{\mu} \psi_{{\boldsymbol k}}$ and 
$A_{\mu} ({\boldsymbol q}, \omega)\!=\! (\phi_{\boldsymbol q} (\omega) /{c^*} , - {\boldsymbol a}_{\boldsymbol q} (\omega))$, respectively.
As shown in Table~\ref{tab:1}, the coupling constant $e_0$ ($>0$) is equal to 
the elementary charge $e$ in the present case.

The Hamiltonian $H + H'(t)$ is, in fact, similar to that of QED. 
We are, however, treating an electromagnetic field with classical theory
while employing the quantum theory of electrons. 
With this treatment, the effects of a mutual Coulomb interaction $e_0^2/\varepsilon_0 q^2$ 
are included through Maxwell's equations in matter as follows:

The electric field 
${\boldsymbol E}_{\boldsymbol q}(\omega)\!=\!- i {\boldsymbol q} \phi_{\boldsymbol q} (\omega)
+ i \omega {\boldsymbol a}_{\boldsymbol q} (\omega)$ and 
the magnetic induction ${\boldsymbol B}_{\boldsymbol q}(\omega) \!=\!i {\boldsymbol q} {\boldsymbol \times} 
{\boldsymbol a}_{\boldsymbol q} (\omega)$ induce electric charge density modulations 
$- e_0 \delta {\rho}_{\boldsymbol q} (\omega)
\!=\! - i {\boldsymbol q} \,{\boldsymbol \cdot} {\boldsymbol P}_{\boldsymbol q} (\omega) $ 
and an electric current $- e_0 \delta {\boldsymbol j}_{\boldsymbol q} (\omega)
\!=\!i {\boldsymbol q} {\boldsymbol \times} {\boldsymbol M}_{\boldsymbol q} (\omega)
\!-\!i \omega{\boldsymbol P}_{\boldsymbol q} (\omega)$ 
with a polarization ${\boldsymbol P}_{\boldsymbol q} (\omega) 
\!=\! \varepsilon_0 \chi_{\rm e}(q,\omega) {\boldsymbol E}_{\boldsymbol q} (\omega)$ and 
a magnetization ${\boldsymbol M}_{\boldsymbol q} (\omega) 
\!=\! \varepsilon_0 c^2 \chi_{\rm m}(q,\omega) {\boldsymbol B}_{\boldsymbol q} (\omega)$~\cite{Note1}.  
The induced electric charge and current can be then written in terms of the electromagnetic potentials as
\begin{align}
- e_0 \delta {\rho}_{\boldsymbol q} (\omega) 
=& - \! \varepsilon_0 \chi_{\rm e}(q,\omega)
\left[ q^2  \phi_{\boldsymbol q} (\omega)
\!-\! \omega \, {\boldsymbol q}  \,{\boldsymbol \cdot}\, {\boldsymbol a}_{\boldsymbol q} (\omega) \right],
\label{eq: 2}
\\
- e_0 \delta {{\boldsymbol j}}_{\boldsymbol q} (\omega) 
=&
- \! \varepsilon_0 \chi_{\rm e}(q,\omega) 
\left[
{\boldsymbol q} \, \omega \phi_{\boldsymbol q} (\omega)
\!-\! \omega^2 {\boldsymbol a}_{\boldsymbol q} (\omega)
\right]
\nonumber
\\
&+  \varepsilon_0 c^2 \chi_{\rm m}(q,\omega)
\left[
q^2 {\boldsymbol a}_{\boldsymbol q} (\omega)
\!-\! {\boldsymbol q} \, {\boldsymbol q} \,{\boldsymbol \cdot}\, {\boldsymbol a}_{\boldsymbol q} (\omega)
\right] ,
\label{eq: 3}
\end{align} 
where $q^2 \!=\! {\boldsymbol q}^2 \!=\! \delta_{ij} q_i q_j$.

Equations~(\ref{eq: 2}) and~(\ref{eq: 3}) enable us to relate the electric and magnetic susceptibilities $\chi_{{\rm e},{\rm m}}(q,\omega)$ to  
the polarization tensor $\Pi_{\rm R}^{\mu\nu}({\boldsymbol q}, {\omega})$, 
which gives the dynamical four-current 
$\delta {J}^{\mu} ({\boldsymbol q},\omega) \!=\! ( {c^*} \! \delta \rho_{\boldsymbol q}(\omega), \delta {\boldsymbol j}_{\boldsymbol q}(\omega) )$ as 
$- e_0 \delta  {J}^{\mu} ({\boldsymbol q},\omega) \!=\! - \varepsilon_0 {c^*}^2  
\Pi_{\rm R}^{\mu\nu}({\boldsymbol q}, {\omega}) A_\nu ({\boldsymbol q},\omega)$.
When comparing the previous equation with Eqs.~(\ref{eq: 2}) and~(\ref{eq: 3}), 
we can write $\Pi_{\rm R}^{\mu\nu}({\boldsymbol q}, {\omega})$ as
\begin{align}
\Pi_{\rm R}^{\mu \nu}({\boldsymbol q}, {\omega}) 
= &
\Big( {\tilde Q}^2 {\tilde g}^{\mu\nu} \!- {\tilde Q}^{\mu} {\tilde Q}^{\nu} \Big)
\left[ \chi_{\rm e}(q,\omega) + \frac{c^2}{{c^*}^2} \chi_{\rm m}(q,\omega) \right]
\nonumber
\\
&- \Big( Q^2 g^{\mu\nu} \!- Q^{\mu} Q^{\nu} \Big) \, \chi_{\rm e}(q,\omega),
\label{eq: 4}
\end{align}
where ${\tilde g}^{\mu\nu} \!=\! {\rm diag}(0,-1,-1,-1)$, 
${\tilde Q}^{\mu} \!=\! (0, {\boldsymbol q})$, 
${\tilde Q}^2 \!=\! -q^2$, 
$g^{\mu\nu} \!=\! {\rm diag}(1,-1,-1,-1)$, 
$Q^{\mu} \!=\! (\omega/{c^*}, {\boldsymbol q})$, and 
$Q^2 \!=\! \omega^2/{c^*}^2 \!- q^2$. 
Equation~(\ref{eq: 4}) has a general form that satisfies the charge conservation 
$Q_{\mu}\Pi_{\rm R}^{\mu \nu}({\boldsymbol q}, {\omega}) = 0$ 
and gauge invariance $Q_{\nu}\Pi_{\rm R}^{\mu \nu}({\boldsymbol q}, {\omega}) = 0$, 
where $Q_{\mu} = (\omega/{c^*}, -{\boldsymbol q})$.

With $\mu\!=\!\nu\!=\!0$ in Eq.~(\ref{eq: 4}), 
we obtain a standard relationship between the electric susceptibility 
and the polarization function as 
$\chi_{\rm e}(q,\omega) \!=\! \Pi_{\rm R}^{00}({\boldsymbol q}, {\omega})/q^2$. 
Multiplying both sides of Eq.~(\ref{eq: 4}) by $g_{\mu\nu} \!=\! g^{\mu\nu}$ and
taking the summation with respect to the repeated Greek indexes, 
we obtain a useful formula for the magnetic susceptibility as
\begin{align}
\chi_{\rm m}(q,\omega) &\!=\! 
- \frac{{c^*}^2}{c^2} \big[
\chi_{\rm e}(q,\omega) + \Delta \chi(q,\omega)
\big],
\label{eq: 5}
\\
\Delta \chi(q,\omega) &\!=\! 
\frac{3Q^2 \chi_{\rm e}(q,\omega) + 
g_{\mu\nu} \Pi_{\rm R}^{\mu\nu}({\boldsymbol q}, {\omega})
}{2 q^2} .
\label{eq: 6}
\end{align}

Because the polarization tensor can be expressed by the Kubo formula~\cite{Kubo57}, 
we can now make microscopic calculations based on the Dirac Hamiltonian, 
Eq.~(\ref{eq: 1}),  
not only for $\chi_{\rm e}(q,\omega)$ 
but also for $\chi_{\rm m}(q,\omega)$ with the use of Eqs.~(\ref{eq: 5}) and (\ref{eq: 6}). 
The detailed calculations are presented in the Supplemental Material~\cite{Supple}, 
where the standard thermal Green function technique 
for nonrelativistic electron gas~\cite{Vignale05} 
is extended to our ``covariant'' electron--hole gas. 
The presented method of calculations is marginally different from that used in QED 
at finite temperatures and densities~\cite{Bechler81}
in that we use an integral representation of the thermal 
Feynman propagator as an artifice [Eq.~(S.3) in Sect. 2 of Ref. \citen{Supple}]. 
We believe that this makes the calculation process 
clearer in the field of condensed matter science.

We first show our results of the imaginary parts of $\chi_{\rm e}(q, |\omega|)$ 
and $\Delta\chi(q, |\omega|)$ for the Dirac electron system
at zero temperature with an arbitrary value of the chemical potential $\mu$ as follows: 
\hfill
\begin{strip}
\rule[3em]{\dimexpr(0.5\textwidth-0.5\columnsep)}{0.4pt}
\begin{align}
{\rm Im} \chi_{\rm e}(q, |\omega|) &= \frac{e_0^2}{16 \pi \varepsilon_0 \hbar {c^*}} \!\int_{-\infty}^{\infty} d x  \left( x^2 \!\!-\! 1 \right) \left[\theta  \left(-Q^2\right)   \theta  \left( x^2 \!\!-\! a^2 \right) - \theta  \left( Q^2 \!\!-\! \frac{4 {m^*}^2 {c^*}^2}{\hbar^2} \right) \theta \left( a^2 \!\!-\! x^2 \right) \right] 
\theta \,\Big( \! \left( x\!-\!b_- \right) \left( b_+ \!-\! x \right) \!\Big) , \label{eq: 7} 
\\ 
{\rm Im} \Delta\chi(q, |\omega|) &= \frac{e_0^2 Q^2}{32 \pi \varepsilon_0 \hbar {c^*} \! q^2} \! \int_{-\infty}^{\infty} d x \left( 3 x^2 \!\!-\! a^2 \right) \left[ \theta \left(-Q^2\right) \theta \left( x^2 \!\!-\! a^2 \right) - \theta \left( Q^2 \!\!-\! \frac{4 {m^*}^2 {c^*}^2}{\hbar^2} \right) \theta \left( a^2 \!\!-\! x^2 \right) \right]  
\theta \,\Big( \! \left( x\!-\!b_- \right) \left( b_+ \!-\! x \right) \!\Big), \label{eq: 8}
\end{align}
\hfill
\rule[-3em]{\dimexpr(0.5\textwidth-0.5\columnsep)}{0.4pt}
\end{strip}
where $a \!\equiv\! \sqrt{1 \!-\! \frac{4{m^*}^2 {c^*}^2}{\hbar^2 Q^2}}$,
$b_\pm  \!\equiv\! \frac{2|\mu| \pm \hbar |\omega|}{\hbar {c^*}q}$, 
and $\theta(x)$ is the Heaviside step function 
(see the derivation in Sects. 3--5 of Ref. \citen{Supple}).
The first terms with $\theta(-Q^2)$ correspond to 
the contributions from intraband electron excitations, 
while the second terms with $\theta(Q^2-4{m^*}^2 {c^*}^2 /\hbar^2)$ 
correspond to the contributions from virtual electron--hole pairs 
excited across the band gap $E_{\rm G} \!=\! 2{m^*} {c^*}^2$. 
Hence, they represent interband effects. 
We note that $a\!>\!0$, $b_+\!>\!0$, and $b_+\!>\!b_-$ by their definitions and 
whether $a \!>\! |b_\pm|$ or $a \!<\! |b_\pm|$ can be determined from the identity
\begin{align}
a^2 \!- b_{\pm}^2 
= \frac{ 4 {c^*}^2 q^2 \left( \mu^2 \!- {m^*}^2 {c^*}^4 \right) -
\left( \hbar {c^*}^2 Q^2 \! \pm 2 |\mu| |\omega| \right)^2 
}{\hbar^2 {c^*}^4 q^2 Q^2} .
\label{eq: 9}
\end{align}
The imaginary part of $\chi_{\rm m}(q, |\omega|)$ 
is then obtained immediately from Eq.~(\ref{eq: 5}).

The complex susceptibilities $\chi_{{\rm e},{\rm m}}(q,\omega)$ 
can be derived from ${\rm Im} \chi_{{\rm e},{\rm m}}(q,\omega)$ 
using the Kramers--Kronig relation.
The Dirac electron system in solids has a natural bandwidth cutoff $E_{\rm \Lambda}$ 
that is caused by the upper limit of energy, and the dispersion of electrons in a solid 
is regarded as a Dirac dispersion when the energy is below this limit.
We therefore define $\chi_{{\rm e},{\rm m}}(q, \omega)$ as contributions from a 
Dirac dispersion and a part of the total susceptibility of the solid. 
Then, the Kramers--Kronig relation leads to
\begin{align}
\chi_{{\rm e},{\rm m}}(q,\omega)
&= - \frac{1}{\pi} \int_{-2 E_\Lambda/\hbar}^{2 E_\Lambda/\hbar} 
d \omega' \,
\frac{ 
{\rm Im} \chi_{{\rm e},{\rm m}}(q,\omega')
}{\omega_+ - \omega'} ,
\label{eq: 10}
\end{align}
where ${\rm Im} \chi_{{\rm e},{\rm m}}(q,\omega') \!=\! 
{\rm sgn}(\omega') {\rm Im} \chi_{{\rm e},{\rm m}}(q,|\omega'|)$ 
and $\omega_+ \!=\! \omega \!+\! i \eta$ 
with $\eta$ being a positive infinitesimal value. 
It is to be noted that, while the imaginary part of the total susceptibility 
is properly estimated by the present Dirac Hamiltonian, 
i.e., by ${\rm Im} \chi_{{\rm e},{\rm m}}(q,\omega)$ for low energies, 
the real part of the total susceptibility can have extra background contributions 
from higher energy regions, which have a weak dependence on $\omega$. 
However, the singular $\omega$ dependence of the real part of the total susceptibility 
for low energies is correctly described by 
${\rm Re} \chi_{{\rm e},{\rm m}}(q,\omega)$ defined in Eq.~(\ref{eq: 10}).

For a finite temperature $T$, 
the susceptibility can be expressed as an integral of the zero-temperature susceptibility 
with respect to the chemical potential~\cite{Maldague78}. By denoting them as 
$\chi_{{\rm e},{\rm m}}(q, \omega; T, \mu)$ 
to show their $T$ and $\mu$ dependences explicitly, 
the finite-temperature electric and magnetic susceptibilities are given by 
(Sect. 6 of Ref. \citen{Supple}) 
\begin{align}
\chi_{\rm e}(q, \omega; T, \mu)
=&
\int_{-\infty}^{\infty} \! d \mu' \,
\frac{\chi_{\rm e}(q, \omega;0,\mu') 
}{4 k_{\rm B}T \cosh^2 \! \frac{\mu - \mu'}{2k_{\rm B}T}} ,
\label{eq: 11}
\\
\chi_{\rm m}(q, \omega; T, \mu)
=& - \frac{{c^*}^2}{c^2} \chi_{\rm e}(q, \omega; T, \mu)
\nonumber
\\
& - \frac{{c^*}^2}{c^2} \int_{-\infty}^{\infty} \! d \mu' \,
\frac{\Delta \chi(q, \omega; 0, \mu') 
}{4 k_{\rm B}T \cosh^2 \! \frac{\mu - \mu'}{2k_{\rm B}T}} .
\label{eq: 12}
\end{align}
Using Eq.~(\ref{eq: 12}), 
the $T$ dependence of the nuclear spin relaxation time for the Dirac electron system 
has recently been calculated~\cite{Hirosawa17}.

In the following, we concentrate on narrow-gap insulators at $T\!=\!0$ 
in which the chemical potential is in the band gap, i.e., $|\mu| \!<\! {m^*} {c^*}^2$. 
For $Q^2 \!<\! 0$ and $|\mu| \!<\! {m^*} {c^*}^2$, Eq.~(\ref{eq: 9}) leads to 
the constraint of $-a \!<\! b_- \!<\! b_+ \!<\! a$. 
Therefore, the first terms corresponding to the intraband contributions 
vanish in Eqs.~(\ref{eq: 7}) and (\ref{eq: 8}). 
For $Q^2 \!>\! 4 {m^*}^2 {c^*}^2/\hbar^2$ and $|\mu| \!<\! {m^*} {c^*}^2$, 
where $b_- \!<\! 0$, 
Eq.~(\ref{eq: 9}) leads to $ b_- \!<\! -a \!<\! a \!<\! b_+$. 
Thus, the second terms corresponding to the interband contributions 
reduce to the integrals calculated from $-a$ to $a$. 
However, because $\int_{-a}^{a} dx\, (3x^2 - a^2) \!=\! 0$,  
we find that ${\rm Im} \Delta\chi(q, |\omega|)$ vanishes. 
By performing the integration $\int_{-a}^{a} dx\, (x^2 - 1)$ 
for ${\rm Im} \chi_{\rm e}(q, |\omega|)$ and using Eq.~(\ref{eq: 5}), we obtain
\begin{align}
{\rm Im} \chi_{\rm e}(q, |\omega|)
&= - \frac{c^2}{{c^*}^2} {\rm Im} \chi_{\rm m}(q, |\omega|)
\nonumber
\\
&= \frac{e_0^2}{24 \pi \varepsilon_0 \hbar {c^*}}
\theta \left(Q^2 \!\!-\! \frac{4 {m^*}^2 {c^*}^2}{\hbar^2} \right) 
a \left( 3 \!-\! a^2 \right) .
\label{eq: 13}
\end{align}
Because $a$ is a function of only $Q^2$, 
the imaginary parts of $\chi_{{\rm e},{\rm m}}(q, |\omega|)$ 
depend on $q$ and $\omega$ only through 
$Q^2 \!=\! \omega^2/{c^*}^2 \!\!-\! q^2$.

The substitution of Eq.~(\ref{eq: 13}) into Eq.~(\ref{eq: 10}) 
for $\hbar {c^*} q \ll E_{\Lambda}$ yields
\begin{align}
\chi_{\rm e}(q,\omega)
&= 
- \frac{c^2}{{c^*}^2}
\chi_{\rm m}(q,\omega)
= - \Pi_2 ( Q_+^2 ) , 
\label{eq: 14}
\end{align}
where $\Pi_2 ( Q_+^2 ) \equiv -\chi_{\rm e}(q,\omega)$ is a function of 
$Q_+^2 \!=\! \omega_+^2/{c^*}^2 \!\!-\! q^2$. 
The presence of the factor $c^2/{c^*}^2$ is caused by the difference in 
the effective Lorentz covariance of the Dirac equation for electrons in solids and
the true Lorentz covariance of Maxwell's equations. 
In fact, if ${c^*}$ is replaced by $c$, Eq.~(\ref{eq: 14}) reduces to 
$\varepsilon_{\rm r} (q, \omega) \mu_{\rm r} (q, \omega) 
= [1 + \chi_{\rm e}(q,\omega)]/[1 - \chi_{\rm m}(q,\omega)]
= 1$ in accordance with the full Lorentz covariance.
From Eq.~(\ref{eq: 14}), $\chi_{\rm m}(q,\omega)$ is opposite in sign to 
$\chi_{\rm e}(q,\omega)$. 
The magnitude of $\chi_{\rm m}(q,\omega)$ is much smaller than that of $\chi_{\rm e}(q,\omega)$ 
by the factor of ${c^*}^2/c^2 \sim 10^{-6}$ in solids. 
Although, the consideration of the anisotropy effects 
is necessary for an improved quantitative evaluation 
as exemplified elsewhere.

Carrying out the integration in Eq.~(\ref{eq: 10}) with Eq.~(\ref{eq: 13}), 
we obtain an explicit expression of $\Pi_2 ( Q_+^2 )$ 
for ${m^*} {c^*}^2\!\ll\!E_{\Lambda}$ as (Sect. 7 of Ref. \citen{Supple}) 
\begin{align}
\Pi_2 ( Q_+^2 )
&= -  \frac{e_0^2}{12 \pi \varepsilon_0 \hbar {c^*}} 
\left[ 
\frac{1}{\pi}\log \frac{M_{\Lambda}^2}{{m^*}^2}
+ P_2 \!\left( \frac{\hbar^2 Q_+^2}{{m^*}^2{c^*}^2} \right) 
\right] ,
\label{eq: 15}
\end{align}
where $M_{\Lambda} \!=\! 2 e^{-5/6} E_\Lambda/{c^*}^2$ 
and $P_2 (s)$ is an analytic function of a complex variable $s$ 
given by
\begin{align}
P_2 (s) =
-\frac{1}{3 \pi} 
+ \frac{1}{\pi}\left( 1 \!+\! \frac{2}{s} \right) 
\left[ 2 \!-\! \sqrt{1 \!-\! \frac{4}{s}}
\log \! \frac{\sqrt{1\!-\!\frac{4}{s}} + 1}{\sqrt{1\!-\!\frac{4}{s}} - 1}
\right] .
\label{eq: 16}
\end{align}
By a series expansion with respect to $s$, 
we can check that $P_2 (s)$ vanishes at $s\!=\!0$. 
From Eqs.~(\ref{eq: 14})--(\ref{eq: 16}), 
we see that
${\rm Re} \chi_{{\rm e},{\rm m}}(q,\omega)$ has a cusp singularity  
at $\omega^2 =  {c^*}^2 q^2 + 4 {m^*}^2 {c^*}^4/\hbar^2$ 
associated with the interband excitations across the band gap.
In QED (see Table~\ref{tab:1}), 
Eq.~(\ref{eq: 15}) corresponds to a well-known result 
for the bare vacuum polarization function, 
and $Z_3 [ \Pi_2 ( Q_+^2 ) - \Pi_2 ( 0 ) ]$ 
describes the physically observable vacuum polarization function up to 
the second order in renormalized coupling~\cite{Peskin95}.

The relationship between the electric and magnetic susceptibilities, 
Eq.~(\ref{eq: 14}), for a zero-temperature insulator is 
directly linked to the emergence of Lorentz covariance in our electron system. 
This is understood as follows: 
substituting Eq.~(\ref{eq: 14}) into Eq.~(\ref{eq: 4}) yields 
the well-known Lorentz covariant form of the polarization tensor as 
$\Pi_{\rm R}^{\mu \nu}({\boldsymbol q}, {\omega}) \!=\! 
( Q^2 g^{\mu\nu} \!-\! Q^{\mu} Q^{\nu} ) \Pi_2 ( Q_+^2 )$; 
inversely, if the polarization tensor has the above form, 
then Eq.~(\ref{eq: 6}) leads to $\Delta\chi(q,\omega)\!=\!0$ 
and therefore Eq.~(\ref{eq: 14}) as a result. 
However, for nonzero temperatures, 
Eq.~(\ref{eq: 14}) does not exactly hold even for the insulating regime of 
$|\mu(T)| \!<\! {m^*} {c^*}^2$
because the second term in Eq.~(\ref{eq: 12}) has nonzero contributions from 
$\Delta \chi(q, \omega; 0, \mu')$ for $|\mu'| \!>\! {m^*} {c^*}^2$. 
(Explicit expressions for the susceptibilities in the metallic region of 
$|\mu| \!>\! {m^*} {c^*}^2$ will be given in a forthcoming paper.) 
This is similar to the situation in which $\mu_{\rm r} (q, \omega)$ deviates from 
$1/\varepsilon_{\rm r} (q, \omega)$  
for QED with nonzero temperatures, but it is a Lorentz covariant theory~\cite{Weldon82}.

In the uniform and static limit of $q\!=\!\omega\!=\!0$, Eq.~(\ref{eq: 14}) reduces to
\begin{align}
\varepsilon_{\rm r}(0,0) = 1 - \frac{c^2}{{c^*}^2}
\chi_{\rm m}(0,0) = \frac{1}{Z_3} .
\label{eq: 17}
\end{align}
Noting that $e_0\!=\!e$ in solids and the fine-structure constant is given by 
$\frac{e^2}{4 \pi \varepsilon_0 \hbar c} \!\approx\! \frac{1}{137}$, 
we can evaluate $Z_3^{-1} = 1 - \Pi_2 ( 0 )$ to be
\begin{align}
\frac{1}{Z_3} \approx 1 + 
1.55 \times 10^{-3} \, \frac{c}{{c^*}}
\log \frac{E_{\Lambda}}{{m^*} {c^*}^2} ,
\label{eq: 18}
\end{align}
where the bandwidth cutoff $E_{\Lambda}$ is on the order of $1$ eV.
The uniform and static magnetic susceptibility is then given by 
$\chi_{\rm m}(0,0) = - 1.55 \times 10^{-3} ({c^*}/c) \log (E_{\Lambda}/{m^*} {c^*}^2 )$, 
which is equivalent to the previous result for large 
diamagnetism~\cite{Fukuyama70,Fukuyama71,Fuseya14}. 
From Eqs.~(\ref{eq: 17}) and~(\ref{eq: 18}), 
we find not only the large diamagnetism [$\chi_{\rm m}(0,0) \!\to\! -\infty$] 
but also a large enhancement in the permittivity 
[$Z_3 \!\equiv\!  1/\varepsilon_{\rm r}(0,0) \!\to\! 0$] 
for ${m^*} {c^*}^2\!\ll\!E_{\Lambda}$ (${m^*} {c^*}^2/E_{\Lambda}\!\to\!0$). 
The physical interpretation is as follows.

In a zero-temperature insulator, 
virtual electron--hole pairs are created and annihilated dynamically by quantum fluctuations 
forming a charge distribution of size $\sim \! h/{m^*}{c^*}$. 
In the presence of an electromagnetic field,
those electron--hole pairs fluctuate 
on the length scale $\sim\! h{c^*}/E_{\Lambda}$; 
in turn, this change reacts to the field. 
This effect is called the self-energy of an electromagnetic field. 
Thus, in the limit of 
${m^*} {c^*}^2/E_{\Lambda} \!=\! (h{c^*}/E_{\Lambda})/(h/{m^*}{c^*}) \!\to\! 0$, 
the charge distribution behaves as a freely deformable distribution that exhibits 
perfect screening [$Z_3 \!\equiv\!  1/\varepsilon_{\rm r}(0,0) \!\to\! 0$] 
in the presence of an external charge on one hand
and perfect diamagnetism [$\chi_{\rm m}(0,0) \!\to\! -\infty$] 
in the presence of an external magnetic field on the other hand~\cite{Note1}.

In summary, we have studied the electrodynamics of Dirac electrons 
in a narrow-gap system to find a remarkable correlation 
between its dielectric and diamagnetic properties. 
Our findings are described by Eqs.~(\ref{eq: 14}),~(\ref{eq: 17}), and~(\ref{eq: 18}). 
These equations show that both the large diamagnetism 
and a large enhancement of the permittivity result from 
virtual electron--hole pair creations across the small band gap, 
i.e., interband effects associated with an electromagnetic field.

\footnotesize{
\textbf{Acknowledgments} 
We thank the very fruitful discussions with Y. Fuseya, T.  Hirosawa, H. Matsuura, T. Mizoguchi, and N. Okuma. 
This work was supported by a Grant-in-Aid for Scientific Research on ``Multiferroics in Dirac electron materials'' (No.15H02108).}

\end{document}


\maketitle

\section{Preliminaries}

First of all, we introduce our notations. 
We use capital letters of the alphabet for four-vectors as follows:
\[
\begin{array}{lll}
K^{\mu} = (k_0, k_1, k_2, k_3), & K_{\mu} = (k_0, -k_1, -k_2, -k_3), 
& K^2 = K_{\mu} K^{\mu}  = k_0^2 - k^2,
\\
{K'}^{\mu} = (k_0', k_1', k_2', k_3'), & {K}_{\mu}' = (k_0', -k_1', -k_2', -k_3'), 
& {K'}^2 = {K}_{\mu}' {K'}^{\mu}  = {k_0'}^2 - {k'}^2,
\\
Q^{\mu} = (q_0, q_1, q_2, q_3), & Q_{\mu} = (q_0, -q_1, -q_2, -q_3), 
&Q^2 = Q_{\mu} Q^{\mu}  = q_0^2 - q^2,
\end{array}
\]
where $k^2 = {\boldsymbol k}^2$, ${k'}^2 = {{\boldsymbol k}'}^2$, and $q^2 = {\boldsymbol q}^2$ 
with ${\boldsymbol k}\!=\! (k_1, k_2, k_3)$, ${\boldsymbol k}'\!=\! (k_1', k_2', k_3')$, and 
${\boldsymbol q} \!=\! (q_1, q_2, q_3)$, respectively. 
The Feynman slash notation $\kksla$ and 
four-vector scalar product $K \!\cdot\! {K'}$ are defined as
\[
{\kksla} 
= K_{\mu} \gamma^{\mu} = k_0 \gamma^0 \!-\! k_i\gamma^i, 
\quad 
K \!\cdot\! {K'} =  K_{\mu} {K'}^{\mu} 
=  k_0 k_0' \!-\!  k_i k_i'
=  k_0 k_0' \!-\! {\boldsymbol k} \!\cdot\! {\boldsymbol k}' .
\] 
The integrals with respect to four-vectors are given as
\[
\int d^4 K \!=\! \int_{-\infty}^{\infty} d k_0 \int d^3k, \quad 
\int d^4K' \!=\! \int_{-\infty}^{\infty} d k_0' \int d^3k' .
\]

In addition, we note the following two identities, 
which will be used for calculations of the polarization tensor in Sect. 3: 
one is
\[
T \sum_{{\omega_k}} 
\frac{1}{(i{\omega_k} + \mu - {c^*} k_0 ) 
(i{\omega_k} + i {\omega_q} + \mu - {c^*} k'_0 )}
= \frac{f( {c^*} k_0) - f( {c^*} k'_0)}{i {\omega_q} + {c^*} k_0 - {c^*} k'_0} \, , 
\tag{S.1}
\]
where $\omega_k$ ($\omega_q$) is an odd (even) Matsubara frequency 
and $f(x) \!=\! [e^{(x-\mu)/T}\!+\!1]^{-1}$ is the Fermi distribution function~\cite{Vignale05};  
the other is 
\[
{\rm Tr} \left[ 
\gamma^{\mu} \!\left(
{\kksla} 
+\! {m^*}{c^*} \right)
\gamma^{\nu} \!\left(
{\kksla}' 
\!+\! {m^*}{c^*} \right)
\right] 
=
4 \left[ 
K^{\mu} {K'}^{\nu} \!+\! {K'}^{\mu} K^{\nu} 
\!-\! (K \!\cdot\! K' \!-\!  {m^*}^2{c^*}^2) g^{\mu\nu} 
\right] ,
\tag{S.2}
\]
which can be easily proved by using the trace identities for the gamma matrices~\cite{Peskin95}
\[
{\rm Tr} (\gamma^\mu) = 0,
\quad
{\rm Tr} (\gamma^\mu \gamma^\nu ) = 4 g^{\mu\nu} \!,
\quad
{\rm Tr} (\gamma^\mu \gamma^\nu \gamma^\rho) = 0, 
\quad
{\rm Tr} (\gamma^\mu \gamma^\nu \gamma^\rho \gamma^\sigma)
= 4 (g^{\mu\nu} g^{\rho\sigma} \!- g^{\mu\rho} g^{\nu\sigma} \!+ g^{\mu\sigma} g^{\nu\rho} ).
\]

\section{Integral representation of the thermal Feynman propagator}

The thermal Feynman propagator $S_{\rm F} ({\boldsymbol k}, i{\omega_k})$  
is defined as
\[
S_{\rm F} ({\boldsymbol k}, i{\omega_k}) \equiv - \frac{1}{2} 
\int_{-1/T}^{1/T} d \tau 
\langle T_\tau \, e^{i (H - \mu N) \tau} \psi_{\boldsymbol k} \, e^{- i  (H - \mu N) \tau} 
{\bar \psi}_{\boldsymbol k} \rangle 
e^{i {\omega_k} \tau} ,
\]
where the (imaginary) time-ordered product is represented by $T_\tau$
and $\langle \cdots \rangle$ denotes the grand canonical average for 
$H \!-\! \mu N$ with 
$N \!=\! \sum_{\boldsymbol k} {\bar \psi}_{{\boldsymbol k}} \gamma^{0} \psi_{{\boldsymbol k}}$
as the electron number operator. 
Then $S_{\rm F} ({\boldsymbol k}, i{\omega_k})$ is obtained as
\[
S_{\rm F} ({\boldsymbol k}, i{\omega_k}) 
= \left[ (i{\omega_k} \!+\! \mu) \gamma^0 \!-\! {c^*} k_i \gamma^i  \!-\!  {m^*} {c^*}^2 \right]^{-1}
= \frac{(i{\omega_k} \!+\! \mu) \gamma^0 
- {c^*} k_i \gamma^i  +  {m^*} {c^*}^2}{
(i{\omega_k} \!+\! \mu)^2 - {c^*}^2 k^2 - {m^*}^2 {c^*}^4} .
\]
To use the standard thermal Green function technique for a covariant field theory, 
we introduce the following integral representation of the thermal Feynman propagator: 
\begin{align}
S_{\rm F} ({\boldsymbol k}, i{\omega_k}) = \int_{-\infty}^{\infty} d k_0 \,
\frac{
{\kksla} 
+ {m^*}{c^*}}{i{\omega_k} \!+\! \mu - {c^*} k_0} \,
{\rm sgn} (k_0) \, \delta (K^2 \!-\! {m^*}^2 {c^*}^2) .
\tag{S.3}
\end{align}
By using
\begin{align}
\delta (K^2 \!-\! {m^*}^2 {c^*}^2) 
= \frac{1}{2 \sqrt{k^2 \!+\! {m^*}^2 {c^*}^2}} 
\left[ \delta \left( k_0 \!-\! \sqrt{k^2 +{m^*}^2 {c^*}^2} \right) 
+ \delta \left( k_0 \!+\! \sqrt{k^2 + {m^*}^2 {c^*}^2} \right) \right] ,
\nonumber
\end{align}
we can confirm that  the right hand side of Eq.~(S.3) equals 
\begin{align}
&\frac{1}{2 \sqrt{k^2 \!+\! {m^*}^2 {c^*}^2}} \!
\left[
\frac{\sqrt{k^2 \!+\! {m^*}^2 {c^*}^2}\gamma^0 \!-\! k_i\gamma^i \!+\! {m^*}{c^*}
}{i {\omega_k} \!+\! \mu \!-\! {c^*} \! \sqrt{k^2 +{m^*}^2 {c^*}^2}}
- \frac{- \sqrt{k^2 +{m^*}^2 {c^*}^2}\gamma^0 \!-\! k_i\gamma^i \!+\! {m^*}{c^*}}{
i {\omega_k} \!+\! \mu \!+\! {c^*} \sqrt{k^2 \!+\! {m^*}^2 {c^*}^2}}
\right]
\nonumber
\\
&= \frac{(i{\omega_k} \!+\! \mu) \gamma^0 \!-\! {c^*} k_i \gamma^i \!+\!  {m^*} {c^*}^2}{
(i{\omega_k} \!+\! \mu)^2 \!-\!  {c^*}^2 k^2 \!-\! {m^*}^2 {c^*}^4} .
\nonumber
\end{align}

\section{Imaginary part of the polarization tensor}

By applying the Kubo formula~\cite{Kubo57}, 
the (retarded) polarization tensor $\Pi_{\rm R}^{\mu\nu}({\boldsymbol q}, {\omega})$
is given by
\begin{align}
\Pi_{\rm R}^{\mu\nu}({\boldsymbol q}, {\omega}) 
= \frac{i e_0^2}{\varepsilon_0 {c^*}^2} 
\int_0^{\infty} \! dt 
\left\langle \left[ e^{i (H - \mu N)t} J^{\,\mu} ({\boldsymbol q}) e^{- i (H - \mu N) t} , J^{\,\nu} ({-{\boldsymbol q}}) \right] \right\rangle 
e^{i \omega t} .
\nonumber
\end{align}
The retarded polarization tensor $\Pi^{\mu\nu}_{\rm R} ({\boldsymbol q}, \omega)$ 
can be obtained from analytic continuation of the thermal polarization tensor 
$\Pi^{\mu\nu} ({\boldsymbol q}, i{\omega_q})$ defined as  
\begin{align}
\Pi^{\mu\nu} ({\boldsymbol q}, i{\omega_q}) 
&\equiv \frac{e_0^2}{\varepsilon_0 {c^*}^2} \int_0^{1/T} d \tau 
\langle  e^{i (H - \mu N)\tau} J^{\,\mu} ({\boldsymbol q}) 
e^{-i (H - \mu N) \tau} 
J^{\,\nu} ({-{\boldsymbol q}}) \rangle 
e^{i \omega_q \tau} .
\nonumber
\end{align}
Then Wick's theorem leads to
\begin{align}
\Pi^{\mu\nu} ({\boldsymbol q}, i{\omega_q}) 
&= - \frac{e_0^2}{\varepsilon_0}\frac{T}{(2\pi)^3} \sum_{{\omega_k}} 
\int d^3k 
{\rm Tr} \left[ 
\gamma^{\mu} S_{\rm F} ({\boldsymbol k}, i {\omega_k}) 
\gamma^{\nu} S_{\rm F} ({\boldsymbol k} \!+\! {\boldsymbol q}, i {\omega_k} \!+\! i {\omega_q}) 
\right] .
\nonumber
\end{align}
The substitution of  Eq.~(S.3) into this equation yields
\begin{align}
\Pi^{\mu\nu} ({\boldsymbol q}, i{\omega_q}) 
&= - \frac{e_0^2}{\varepsilon_0}\frac{T}{(2\pi)^3} \sum_{{\omega_k}} 
\iint d^3k d^3k'
{\rm Tr} \left[ 
\gamma^{\mu} S_{\rm F} ({\boldsymbol k}, i {\omega_k}) 
\gamma^{\nu} S_{\rm F} ({\boldsymbol k}', i {\omega_k} \!+\! i {\omega_q}) 
\right]
\delta^3 ({\boldsymbol k} \!-\! {\boldsymbol k}' \!+\! {\boldsymbol q}) 
\nonumber
\\
&= - \frac{e_0^2}{\varepsilon_0} \iint \frac{d^3k d^3k'}{(2\pi)^3} 
\! 
\int_{-\infty}^{\infty} \! dk_0\, 
{\rm sgn} (k_0) \, \delta (K^2 \!-\! {m^*}^2 {c^*}^2) \,
\! 
\int_{-\infty}^{\infty} \! dk_0'\,
{\rm sgn} (k_0') \, \delta ({K'}^2 \!-\! {m^*}^2 {c^*}^2)
\nonumber
\\
&\quad \times
T \sum_{{\omega_k}} 
\frac{
{\rm Tr} \left[ 
\gamma^{\mu} \!\left(
{\kksla} 
+ {m^*}{c^*} \right)
\gamma^{\nu} \!\left(
{\kksla}' 
+ {m^*}{c^*} \right)
\right]
}{(i{\omega_k} + \mu - {c^*} k_0) 
(i{\omega_k} + i {\omega_q} + \mu - {c^*} k'_0)} \,
\delta^3 ({\boldsymbol k} \!-\! {\boldsymbol k}' \!+\! {\boldsymbol q}) .
\nonumber
\end{align}
Using Eqs.~(S.1) and (S.2),
we obtain $\Pi^{\mu\nu} ({\boldsymbol q}, i{\omega_q})$ as
\begin{align}
\Pi^{\mu\nu} ({\boldsymbol q}, i{\omega_q}) 
&= - \frac{e_0^2}{2\pi^{3}\varepsilon_0} 
\int d^4 K \,
{\rm sgn} (k_0) \, \delta (K^2 \!-\! {m^*}^2 {c^*}^2) \,
\int d^4K' \,
{\rm sgn} (k_0') \, \delta ({K'}^2 \!-\! {m^*}^2 {c^*}^2)
\nonumber
\\
&\quad \times
\frac{
K^{\mu} {K'}^{\nu} \!+\! {K'}^{\mu} K^{\nu} 
\!-\! (K \!\cdot\! K' \!-\!  {m^*}^2{c^*}^2) g^{\mu\nu} 
}{
i {\omega_q} + {c^*} k_0 - {c^*} k'_0
}
\left[ f( {c^*} k_0) - f( {c^*} k'_0) \right]
\delta^3 ({\boldsymbol k} - {\boldsymbol k}' + {\boldsymbol q}) .
\nonumber
\end{align}

Now we perform the analytic continuation of $\Pi^{\mu\nu}_{\rm R} ({\boldsymbol q}, \omega) 
\!=\! \Pi^{\mu\nu} ({\boldsymbol q}, i{\omega_q})|_{i{\omega_q}\to\omega+i\eta}$ 
with $\eta$ being a positive infinitesimal value.  
The imaginary part of $\Pi^{\mu\nu}_{\rm R} ({\boldsymbol q}, \omega)$ is  
then given by
\begin{align}
{\rm Im} \Pi^{\mu\nu}_{\rm R} ({\boldsymbol q}, \omega) 
&= \frac{e_0^2}{2\pi^2 \varepsilon_0 {c^*}} 
\int d^4 K \,
{\rm sgn} (k_0) \, \delta (K^2 \!-\! {m^*}^2 {c^*}^2) \,
\int d^4K' \,
{\rm sgn} (k_0') \, \delta ({K'}^2 \!-\! {m^*}^2 {c^*}^2)
\nonumber
\\
&\quad \times
\left[ 
K^{\mu} {K'}^{\nu} \!+\! {K'}^{\mu} K^{\nu} 
\!-\! (K \!\cdot\! K' \!-\!  {m^*}^2{c^*}^2) g^{\mu\nu} 
\right]
\left[ f( {c^*} k_0) - f( {c^*} k'_0) \right]
\delta^4 (K\!-\! K' \!+\! Q) ,
\nonumber
\end{align}
where $\delta^4  (K\!-\! K' \!+\! Q) 
= \delta (k_0 \!-\! k_0' \!+\! q_0)\, 
\delta^3 ({\boldsymbol k} \!-\! {\boldsymbol k}' \!+\! {\boldsymbol q})$
with $q_0 \!=\! \omega/{c^*}$.  
Due to the presence of the delta functions, we can replace 
$K^{\mu} {K'}^{\nu} \!+ {K'}^{\mu} K^{\nu}$ and $K \!\cdot\! K'$
by $K^{\mu} K^{\nu} \!+ {K'}^{\mu} {K'}^{\nu} - Q^{\mu} Q^{\nu}$
and ${m^*}^2{c^*}^2 \!- Q^2/2$, respectively. 
This replacement leads to
\begin{align}
{\rm Im} \Pi^{\mu\nu}_{\rm R} ({\boldsymbol q}, \omega) 
&= \frac{e_0^2}{4\pi^2 \varepsilon_0 {c^*}} 
\int d^4 K \,
{\rm sgn} (k_0) \, \delta (K^2 \!-\! {m^*}^2 {c^*}^2) \,
\int d^4K' \,
{\rm sgn} (k_0') \, \delta ({K'}^2 \!-\! {m^*}^2 {c^*}^2)
\nonumber
\\
&\quad \times
\left( 
2 K^{\mu} K^{\nu} \!+ 2 {K'}^{\mu} {K'}^{\nu} \!- 2 Q^{\mu} {Q}^{\nu}
\!+\!  Q^2 g^{\mu\nu} 
\right)
\left[ f( {c^*} k_0) - f( {c^*} k'_0) \right]
\delta^4 (K\!-\! K' \!+\! Q) .
\tag{S.4}
\end{align}

\section{Imaginary parts of $\chi_{\rm e}(q,\omega)$ and $\Delta \chi(q,\omega)$}

From Eq.~(S.4),
we obtain the following expressions for the imaginary parts of 
$\chi_{\rm e}(q,\omega) = \Pi_{\rm R}^{00}({\boldsymbol q}, {\omega})/q^2$
and
$\Delta \chi(q,\omega) = \left[ 3Q^2 \chi_{\rm e}(q,\omega) + 
g_{\mu\nu} \Pi_{\rm R}^{\mu\nu}({\boldsymbol q}, {\omega})\right]/2 q^2$: 
\begin{align}
{\rm Im} \chi_{\rm e}(q, \omega)
&= \frac{e_0^2}{4\pi^2\varepsilon_0 {c^*} q^2} 
\int d^4 K \,
{\rm sgn} (k_0) \, \delta (K^2 \!-\! {m^*}^2 {c^*}^2) \,
\int d^4K' \,
{\rm sgn} (k_0') \, \delta ({K'}^2 \!-\! {m^*}^2 {c^*}^2)
\nonumber
\\
&\quad \times
\left( 
2k_0^2 + 2k_0^{\prime2} - 2q_0^2 + Q^2
\right)
\left[ f( {c^*} k_0) - f( {c^*} k'_0) \right]
\delta^4 (K\!-\! K' \!+\! Q) ,
\tag{S.5}
\\
{\rm Im} \Delta\chi(q, \omega)
&= \frac{e_0^2}{8\pi^2 \varepsilon_0 {c^*} q^4} 
\int d^4 K \,
{\rm sgn} (k_0) \, \delta (K^2 \!-\! {m^*}^2 {c^*}^2) \,
\int d^4K' \,
{\rm sgn} (k_0') \, \delta ({K'}^2 \!-\! {m^*}^2 {c^*}^2)
\nonumber
\\
&\quad \times
\left[ 
3 Q^2 \left( 2k_0^2 + 2k_0^{\prime2} - 2q_0^2 + Q^2 \right)
+ 2 q^2\left( 2 {m^*}^2 {c^*}^2 \!+  Q^2\right)
\right]
\nonumber
\\
&\quad \times
\left[ f( {c^*} k_0) - f( {c^*} k'_0) \right]
\delta^4 (K\!-\! K' \!+\! Q) .
\tag{S.6}
\end{align}
Noting that the integrands in Eqs.~(S.5) and~(S.6) do not depend 
on ${\boldsymbol k}$ and ${\boldsymbol k}'$ except in the delta functions, 
we first consider the integral
\begin{align}
\iint d^3 k d^3 k' 
\delta \left( K^2 \!-\! {m^*}^2 {c^*}^2 \right) 
\delta \left( {K}^{\prime2} \!-\! {m^*}^2 {c^*}^2 \right) 
\delta^3 \! \left( {\boldsymbol k} \!-\! {\boldsymbol k}' \!+\! {\boldsymbol q} \right) .
\nonumber
\end{align}
This integral equals
\begin{align}
&\int d^3 k \,
\delta \left( k_0^2\!-\! k^2 \!-\! {m^*}^2 {c^*}^2 \right) 
\delta \left( k_0^{\prime2}\!-\! ({\boldsymbol k} 
\!+\! {\boldsymbol q} )^2 \!-\! {m^*}^2 {c^*}^2 \right)
\nonumber
\\
&= 
\frac{\pi}{2 q} 
\int_0^{\infty} \! d k^2 
\delta \left( k^2\!-\! k_0^2 +\! {m^*}^2 {c^*}^2 \right) 
\theta \left( k^2  \!-\! \frac{(k_0^{\prime2} \!-\! k_0^2 \!-\! q^2)^2}{4 q^2} \right)
\nonumber
\\
&= 
\frac{\pi}{2 q} \,
\theta \left( 4 q^2 (k_0^2 \!-\! {m^*}^2 {c^*}^2) \!-\! (k_0^{\prime2} \!-\! k_0^2 \!-\! q^2)^2 \right) .
\nonumber
\end{align}
Then Eq.~(S.5) leads to
\begin{align}
{\rm Im} \chi_{\rm e}(q, \omega)
&= \frac{e_0^2}{4\pi^2 {c^*} \varepsilon_0 q^2} 
\int_{-\infty}^{\infty} d k_0 \,
{\rm sgn} (k_0) 
\int_{-\infty}^{\infty}  d k_0' \,
{\rm sgn} (k_0') 
\left( 
2k_0^2 \!+\! 2k_0^{\prime2} \!-\! 2q_0^2 \!+\! Q^2
\right)
\nonumber
\\
&\quad \times
\frac{\pi}{2 q} \,
\theta \left( 4 q^2 (k_0^2 \!-\! {m^*}^2 {c^*}^2) \!-\! (k_0^{\prime2} \!-\! k_0^2 \!-\! q^2)^2 \right)
\left[ f( {c^*} k_0) \!-\! f( {c^*} k'_0) \right]
\delta (k_0\!-\! k_0' \!+\! q_0) ,
\nonumber
\\
&= \frac{e_0^2}{4\pi^2 {c^*} \varepsilon_0 q^2} 
\int_{-\infty}^{\infty}  d k_0 \,
{\rm sgn} (k_0) \,
{\rm sgn} (k_0\!+\!q_0) 
\left[ \left( 2 k_0 \!+\! q_0 \right)^2 \!-\! q^2 \right]
\nonumber
\\
&\quad \times
\frac{\pi}{2 q} \,
\theta \left( 
- Q^2 ( 2 k_0 \!+\! q_0 )^2 \!+ q^2 ( Q^2 \!-\! 4 {m^*}^2 {c^*}^2 )
\right)
\left[ f( {c^*} k_0) \!-\! f( {c^*} k_0 \!+\! {c^*}\!q_0) \right] .
\nonumber
\end{align}
Changing the integral variable from  $k_0 \!=\! (q x \!-\! q_0)/2$ to $x$
with $q_0 \!=\! \omega/{c^*}$, we get
\begin{align}
{\rm Im} \chi_{\rm e}(q, \omega)
&= \frac{e_0^2}{16 \pi \varepsilon_0 {c^*}}
\int_{-\infty}^{\infty}  \! d x \left( x^2 \!-\! 1 \right)
{\rm sgn} \! \left( x^2 \!-\! \frac{\omega^2}{{c^*}^2 q^2} \right) 
\theta\left( 
- Q^2 x^2 \!+\! Q^2 \!-\! 4 {m^*}^2 {c^*}^2 
\right)
\nonumber
\\
&\quad \times
\left[ f \left(\frac{ {c^*} \! q x \!-\! \omega}{2} \right) 
\!-\! f \left( \frac{ {c^*}\!q x \!+\! \omega}{2} \right) \right] .
\nonumber
\end{align}

In the above equation, the Heaviside step function can be separated into two parts as
\begin{align}
\theta \left( 
- Q^2 x^2 \!+\! Q^2 \!-\! 4 {m^*}^2 {c^*}^2 
\right)
&= \theta \left(-Q^2\right) 
\theta \left( x^2 \!-\! \frac{Q^2 \!\!-\!4{m^*}^2 {c^*}^2}{Q^2} \right)
+ 
\theta \left(Q^2\right) 
\theta \left( \frac{Q^2 \!\!-\!4{m^*}^2 {c^*}^2}{Q^2} \!-\! x^2 \right)
\nonumber
\\
&= 
\theta \left(-Q^2\right) 
\theta \left( x^2 \!-\! a^2 \right)
+ 
\theta \left( Q^2 \!\!-\!4{m^*}^2 {c^*}^2 \right) 
\theta \left( a^2 \!-\! x^2 \right) ,
\nonumber
\end{align}
where $a \!\equiv\! \sqrt{\frac{Q^2 - 4{m^*}^2 {c^*}^2}{Q^2}}$ is 
a positive real function of $Q^2$ in the presence of 
$\theta \left(-Q^2\right)$ or $\theta \left( Q^2 \!\!-\!4{m^*}^2 {c^*}^2 \right)$.
Furthermore, the sign function takes 
$1$ for the constraint imposed by $\theta \left(-Q^2\right) \theta \left( x^2 \!-\! a^2 \right)$ 
and $-1$ for the constraint imposed by  
$\theta \left( Q^2 \!\!-\!4{m^*}^2 {c^*}^2\right) \theta \left( a^2 \!-\! x^2 \right)$.
This can be easily understood when the argument of the sign function is written as
\[
x^2 \!-\! \frac{\omega^2}{{c^*}^2 q^2}
= x^2 \!-\! a^2 \!-\! \frac{4{m^*}^2 {c^*}^2}{Q^2}\!-\! \frac{Q^2}{q^2} .
\]

Then, we obtain the imaginary part of $\chi_{\rm e}(q, \omega)$ as
\begin{align}
{\rm Im} \chi_{\rm e}(q, \omega)
&= \frac{e_0^2}{16 \pi \varepsilon_0 {c^*}}
\theta \left(-Q^2\right)  
\int_{-\infty}^{\infty} \! d x \left( x^2 \!-\! 1 \right) 
\theta \left( x^2 \!-\! a^2 \right)
\left[ f \left( \frac{{c^*} \! q x \!-\! \omega}{2} \right) 
\!-\! f \left( \frac{{c^*}\!q x \!+\! \omega}{2} \right) \right]
\nonumber
\\
&\quad \!-\!  \frac{e_0^2}{16 \pi \varepsilon_0 {c^*}} 
\theta \left(Q^2 \!\!-\! 4 {m^*}^2 {c^*}^2 \right) 
\int_{-\infty}^{\infty} \! d x \left( x^2 \!-\! 1 \right)
\theta \left( a^2 \!-\! x^2 \right)
\left[ f \left( \frac{{c^*} \! q x \!-\! \omega}{2} \right) 
\!-\! f \left( \frac{{c^*}\!q x \!+\! \omega}{2} \right) \right] .
\tag{S.7}
\end{align}
In a similar way from Eq.~(S.6), 
we also obtain the imaginary part of $\Delta\chi(q, \omega)$ as
\begin{align}
{\rm Im} \Delta\chi(q, \omega)
&= \frac{e_0^2}{32 \pi \varepsilon_0 {c^*}} \frac{Q^2}{q^2} \theta \left(-Q^2\right)
\int_{-\infty}^{\infty} \! d x 
\left( 3 x^2 \!-\! a^2 \right) \theta \left( x^2 \!-\! a^2 \right)
\left[ f \left( \frac{{c^*} \! q x \!-\! \omega}{2} \right) 
\!-\! f \left( \frac{{c^*}\!q x \!+\! \omega}{2} \right) \right]
\nonumber
\\
&\quad 
- \frac{e_0^2}{32 \pi \varepsilon_0 {c^*}} \frac{Q^2}{q^2} 
\theta \left(Q^2 \!\!-\! 4 {m^*}^2 {c^*}^2 \right) 
\nonumber
\\
&\quad \times
\int_{-\infty}^{\infty} \! d x 
\left( 3 x^2 \!-\! a^2 \right) \theta \left( a^2 \!-\! x^2 \right)
\left[ f \left( \frac{{c^*} \! q x \!-\! \omega}{2} \right) 
\!-\! f \left( \frac{{c^*}\!q x \!+\! \omega}{2} \right) \right] .
\tag{S.8}
\end{align}
We note that similar equations as Eqs.~(S.7) and (S.8) can be found in the study 
of quantum electrodynamics at finite temperatures and 
densities~\cite{Bechler81}.

At zero temperature, Eqs.~(S.7) and (S.8) reduce to
\begin{align}
{\rm Im} \chi_{\rm e}(q, \omega)
&= \frac{e_0^2}{16 \pi \varepsilon_0 {c^*}}
\theta \left(-Q^2\right)  
\int_{-\infty}^{\infty} \! d x \left( x^2 \!-\! 1 \right) 
\theta \left( x^2 \!-\! a^2 \right)
\left[\theta \left( {\mu} \!-\! \frac{{c^*} \! q x \!-\! \omega}{2} \right) 
\!-\! \theta \left( {\mu} \!-\! \frac{{c^*} \! q x \!+\! \omega}{2} \right) \right]
\nonumber
\\
&\quad \!-\!  \frac{e_0^2}{16 \pi \varepsilon_0 {c^*}} 
\theta \left(Q^2 \!\!-\! 4 {m^*}^2 {c^*}^2 \right) 
\nonumber
\\
&\quad \times
\int_{-\infty}^{\infty} \! d x \left( x^2 \!-\! 1 \right)
\theta \left( a^2 \!-\! x^2 \right)
\left[\theta \left( {\mu} \!-\! \frac{{c^*} \! q x \!-\! \omega}{2} \right) 
\!-\! \theta \left( {\mu} \!-\! \frac{{c^*} \! q x \!+\! \omega}{2} \right) \right] ,
\tag{S.9}
\\
{\rm Im} \Delta\chi(q, \omega)
&= \frac{e_0^2}{32 \pi \varepsilon_0 {c^*}} \frac{Q^2}{q^2} \theta \left(-Q^2\right)
\int_{-\infty}^{\infty} \! d x 
\left( 3 x^2 \!-\! a^2 \right) \theta \left( x^2 \!-\! a^2 \right)
\left[\theta \left( {\mu} \!-\! \frac{{c^*} \! q x \!-\! \omega}{2} \right) 
\!-\! \theta \left( {\mu} \!-\! \frac{{c^*} \! q x \!+\! \omega}{2} \right) \right]
\nonumber
\\
&\quad 
- \frac{e_0^2}{32 \pi \varepsilon_0 {c^*}} \frac{Q^2}{q^2} 
\theta \left(Q^2 \!\!-\! 4 {m^*}^2 {c^*}^2 \right) 
\nonumber
\\
&\quad \times
\int_{-\infty}^{\infty} \! d x 
\left( 3 x^2 \!-\! a^2 \right) \theta \left( a^2 \!-\! x^2 \right)
\left[\theta \left( {\mu} \!-\! \frac{{c^*} \! q x \!-\! \omega}{2} \right) 
\!-\! \theta \left( {\mu} \!-\! \frac{{c^*} \! q x \!+\! \omega}{2} \right) \right] .
\tag{S.10}
\end{align}

\section{Derivation of Eqs.~(7) and (8)}

As seen from Eqs.~(S.7) and (S.8), 
the imaginary parts of $\chi_{\rm e}(q, \omega)$ and $\Delta\chi(q, \omega)$ 
are odd functions of the frequency $\omega$     
while even functions of the chemical potential $\mu$. 
The former is related to the analyticity while the latter to the particle-hole symmetry.
By taking them into account, 
Eqs.~(S.9) and (S.10) can be transformed into
\begin{align}
{\rm Im} \chi_{\rm e}(q, \omega)
&= \frac{e_0^2}{16 \pi \varepsilon_0 {c^*}}
{\rm sgn}(\omega) \,
\theta \left(-Q^2\right)  
\int_{-\infty}^{\infty} \! d x \left( x^2 \!-\! 1 \right) 
\theta \left( x^2 \!-\! a^2 \right)
\Big[\theta \left( b_+ \!-\! x \right) 
\!-\! \theta \left( b_- \!-\! x \right) \Big]
\nonumber
\\
&\quad \!-\!  \frac{e_0^2}{16 \pi \varepsilon_0 {c^*}} 
{\rm sgn}(\omega) \,
\theta \left(Q^2 \!\!-\! 4 {m^*}^2 {c^*}^2 \right) 
\nonumber
\\
&\quad \times
\int_{-\infty}^{\infty} \! d x \left( x^2 \!-\! 1 \right)
\theta \left( a^2 \!-\! x^2 \right)
\Big[\theta \left( b_+ \!-\! x \right) 
\!-\! \theta \left( b_- \!-\! x \right) \Big] ,
\nonumber
\\
{\rm Im} \Delta\chi(q, \omega)
&= \frac{e_0^2}{32 \pi \varepsilon_0 {c^*}} \frac{Q^2}{q^2} 
{\rm sgn}(\omega) \,
\theta \left(-Q^2\right)
\int_{-\infty}^{\infty} \! d x 
\left( 3 x^2 \!-\! a^2 \right) \theta \left( x^2 \!-\! a^2 \right)
\Big[\theta \left( b_+ \!-\! x \right) 
\!-\! \theta \left( b_- \!-\! x \right) \Big]
\nonumber
\\
&\quad 
- \frac{e_0^2}{32 \pi \varepsilon_0 {c^*}} \frac{Q^2}{q^2} 
{\rm sgn}(\omega) \,
\theta \left(Q^2 \!\!-\! 4 {m^*}^2 {c^*}^2 \right) 
\nonumber
\\
&\quad \times
\int_{-\infty}^{\infty} \! d x 
\left( 3 x^2 \!-\! a^2 \right) \theta \left( a^2 \!-\! x^2 \right)
\Big[\theta \left( b_+ \!-\! x \right) 
\!-\! \theta \left( b_- \!-\! x \right) \Big] ,
\nonumber
\end{align}
where $b_\pm \equiv \frac{2|\mu| \pm |\omega|}{{c^*}q}$.
The above equations for ${\rm Im} \chi_{\rm e}(q, \omega)$ and 
${\rm Im} \Delta\chi(q, \omega)$ correspond to Eqs.~(7) and~(8) in the text, 
respectively.

\section{Derivation of Eqs.~(11) and (12)}

For the noninteracting electron gas, it is known that 
the finite-temperature polarizability can be given as an integral 
of the zero-temperature polarizability with respect to the chemical potential~\cite{Maldague78}. 
In the same way as for the electron gas, 
we can express the finite-temperature susceptibility
as an integral of the zero-temperature susceptibility with respect to the chemical potential 
for our ``covariant'' electron-hole gas.

Let us write 
$\chi_{\rm e}(q, \omega)$, $\chi_{\rm m}(q, \omega)$, and $\Delta\chi(q, \omega)$ as
$\chi_{\rm e}(q, \omega; T, \mu)$, $\chi_{\rm m}(q, \omega; T, \mu)$, 
and $\Delta\chi(q, \omega; T, \mu)$, respectively, 
to show their dependences on $T$ and $\mu$ explicitly. 
By using Eqs.~(S.9) and (S.10) with the identity~\cite{Maldague78}
\[
f(x) = 
\frac{1}{e^{(x-\mu)/T} \!+\!1} 
= \int_{-\infty}^{\infty} d \mu' \frac{\theta \left( \mu' \!-\! x \right)
}{4 T \cosh^2 \frac{\mu - \mu'}{2T}} ,
\]
we can express Eqs.~(S.7) and (S.8) as
\begin{align}
{\rm Im} \chi_{{\rm e}}(q, \omega; T, \mu)
&= 
\int_{-\infty}^{\infty} \! d \mu' \,
\frac{{\rm Im} \chi_{{\rm e}}(q, \omega;0,\mu') 
}{4 T \cosh^2 \frac{\mu - \mu'}{2T}}, 
\nonumber
\\
{\rm Im} \Delta\chi(q, \omega; T, \mu)
&= 
\int_{-\infty}^{\infty} \! d \mu' \,
\frac{{\rm Im}\Delta\chi(q, \omega;0,\mu') 
}{4 T \cosh^2 \frac{\mu - \mu'}{2T}} . 
\nonumber
\end{align}
Because their real parts are related to their imaginary parts by the Kramers--Kronig relation, 
the finite-temperature electric and magnetic susceptibilities can be given as
\begin{align}
\chi_{{\rm e}}(q, \omega; T, \mu)
&= \int_{-\infty}^{\infty} \! d \mu' \,
\frac{\chi_{{\rm e}}(q, \omega;0,\mu') 
}{4 T \cosh^2 \frac{\mu - \mu'}{2T}}, 
\nonumber
\\
\chi_{\rm m}(q, \omega; T, \mu)
&= - \frac{{c^*}^2}{c^2} 
\left[ \chi_{\rm e}(q, \omega; T, \mu)
+ \int_{-\infty}^{\infty} \! d \mu' \,
\frac{\Delta \chi(q, \omega; 0, \mu') 
}{4 T \cosh^2 \frac{\mu - \mu'}{2T}} 
\right] .
\nonumber
\end{align}
The above equations for $\chi_{{\rm e}}(q, \omega; T, \mu)$ and 
$\chi_{\rm m}(q, \omega; T, \mu)$ 
correspond to Eqs.~(11) and (12) in the text, respectively.

\section{Derivation of Eq.~(15)}

From Eqs.~(10) and (13) in the text, 
the electric susceptibility $\chi_{{\rm e}}(q,\omega)$ is given 
for $T = 0$ and $|\mu| < {m^*} {c^*}^2$ as
\begin{align}
\chi_{{\rm e}}(q,\omega)
&= - \frac{1}{\pi} \int_{0}^{2 E_\Lambda} 
d \omega' \,
\left( \frac{1}{\omega_+ - \omega'} 
-
\frac{1}{\omega_+ + \omega'} 
\right)
{\rm Im} \chi_{{\rm e}}(q,\omega')
\nonumber
\\
&=
\frac{e_0^2}{24 \pi^2 \varepsilon_0  {c^*}} 
\int_{0}^{2 E_\Lambda} 
d \omega' \,
\frac{2 \omega'}{\omega'^2 - \omega_+^2}
a\left(Q'^2\right) \left[ 3 - a^2 \!\left(Q'^2\right) \right]
\theta \left(Q'^2 \!\!-\! 4 {m^*}^2 {c^*}^2 \right) 
\nonumber
\\
&=
\frac{e_0^2}{24 \pi^2 \varepsilon_0 {c^*}} 
\int_{4 {m^*}^2 {c^*}^2}^{\,4 E_\Lambda^2/{c^*}^2 -\, q^2} 
\!\! \frac{d Q'^2}{Q'^2 - Q_+^2}
a\left(Q'^2\right) \left[ 3 - a^2 \!\left(Q'^2\right) \right] ,
\nonumber
\end{align}
where $a (Q'^2) = \sqrt{1 - 4 {m^*}^2 {c^*}^2/Q'^2}$ 
with $Q'^2 = \omega'^2/ {c^*}^2  - q^2$. 
For $q \ll E_{\Lambda}/{c^*}$, 
the electric susceptibility $\chi_{{\rm e}}(q,\omega)$ 
becomes a function of only $Q_+^2$.
Then we introduce $\Pi_2 \left(Q_+^2\right) \equiv - \chi_{{\rm e}}(q,\omega)$ as
\begin{align}
\Pi_2 \left(Q_+^2 \right)
&= - \frac{e_0^2}{24 \pi^2 \varepsilon_0 {c^*}} 
\int_{4 {m^*}^2 {c^*}^2}^{\,4 E_\Lambda^2/{c^*}^2} 
\!\! \frac{d Q'^2}{Q'^2 - Q_+^2}
a\left(Q'^2\right) \left[ 3 - a^2 \!\left(Q'^2\right) \right] .
\nonumber
\end{align}
Changing the integral variable from $Q'^2$ to 
$a' = a (Q'^2)$, 
we obtain
\begin{align}
\Pi_2 \left(Q_+^2\right)
=& - \frac{e_0^2}{12 \pi^2 \varepsilon_0 {c^*}} 
\int_{0}^{\sqrt{1 - {m^*}^2 {c^*}^4/E_\Lambda^2}} 
da' \left[ \frac{2}{1-a'^2} + \frac{4}{s} - \left(1 + \frac{2}{s} \right)  
\frac{2 \left(1 - \frac{4}{s} \right)}{1 - \frac{4}{s}  - a'^2}
\right] ,
\nonumber
\end{align}
where $s = Q_+^2/{m^*}^2 {c^*}^2$. 
The integral with respect to $a'$ is elementary;
for ${m^*} {c^*}^2/E_\Lambda \ll 1$, it leads to
\begin{align}
\Pi_2 \left(Q_+^2\right)
=& - \frac{e_0^2}{12 \pi^2 \varepsilon_0 {c^*}} 
\left[ \log \frac{4 E_\Lambda^2}{{m^*}^2 {c^*}^4} + \frac{4}{s}
- \left( 1 + \frac{2}{s} \right) \sqrt{1 - \frac{4}{s}}
\log \frac{\sqrt{1-\frac{4}{s}} + 1}{\sqrt{1-\frac{4}{s}} - 1}
\right]  .
\nonumber
\end{align}
The above equation corresponds to Eq.~(15) in the text.